\UseRawInputEncoding
\documentclass[10pt,final,doublecolumn]{IEEEtran}
\usepackage{graphicx,cite,epsfig}
\usepackage{amsmath} 
\usepackage{amssymb}  
\DeclareMathOperator *{\argmin}{argmin}
\usepackage{pifont}
\usepackage{stmaryrd}
\usepackage{bbding}
\usepackage{subfigure}
\usepackage{algorithm}
\usepackage{algorithmic}
\usepackage{array}
\usepackage{amsthm}
\usepackage{multirow}
\usepackage{url}
\usepackage{mathrsfs}
\usepackage{color, soul} 
\usepackage{mathrsfs}
\usepackage{float}

\usepackage{booktabs}
\usepackage{threeparttable}
\usepackage{amssymb}
\IEEEoverridecommandlockouts
\begin{document}
\title{Fast polar codes for terabits-per-second throughput communications}

\author{\IEEEauthorblockN{
Jiajie~Tong,~\IEEEmembership{Member,~˜IEEE}, Xianbin~Wang,~\IEEEmembership{Member,~˜IEEE}, Qifan~Zhang,~\IEEEmembership{Member,~˜IEEE}, Huazi~Zhang,~\IEEEmembership{Member,~˜IEEE}, Rong~Li,~\IEEEmembership{Member,~˜IEEE}, Jun~Wang,~\IEEEmembership{Member,~˜IEEE} and Wen~Tong,~\IEEEmembership{Fellow,~˜IEEE}
\thanks{Part of this paper was presented in an invited talk at the 2021 International Symposium on Information Theory (ISIT).}
\thanks{Jiajie~Tong, Xianbin~Wang, Huazi~Zhang, Rong~Li and Jun~Wang are with Huawei Technologies Co. Ltd., China.}
\thanks{Qifan~Zhang, Huazi~Zhang and Wen~Tong are with Huawei Technologies Canada Co. Ltd., Canada.}}}\maketitle

\begin{abstract}
Targeting high-throughput and low-power communications, we implement two successive cancellation (SC) decoders for polar codes. With $16nm$ ASIC technology, the area efficiency and energy efficiency are $4Tbps/mm^2$ and $0.63pJ/bit$, respectively, for the unrolled decoder, and $561Gbps/mm^2$ and $1.21pJ/bit$, respectively, for the recursive decoder. To achieve such a high throughput, a novel code construction, coined as fast polar codes, is proposed and jointly optimized with a highly-parallel SC decoding architecture. First, we reuse existing modules to fast decode more outer code blocks, and then modify code construction to facilitate faster decoding for all outer code blocks up to a degree of parallelism of $16$. Furthermore, parallel comparison circuits and bit quantization schemes are customized for hardware implementation. Collectively, they contribute to an $2.66\times$ area efficiency improvement and $33\%$ energy saving over the state of the art.

\end{abstract}
\begin{IEEEkeywords}
Fast polar codes, Tbps communication, fast decoding, recursive decoder, unroll decoder.
\end{IEEEkeywords}

\section{Introduction}\label{section_introductions}
\subsection{Motivations and Background}
Higher throughput has always been a primary target along the course of mobile communications evolution. Driven by high data rate applications such as virtual/augmented reality (VR/AR) applications, the sixth generation wireless technology (6G) requires a peak throughput of $1Tbp/s$ \cite{6G}.
This is roughly a $50\times\sim 100\times$ increase over the $10\sim20Gbp/s$ target throughput for 5G standards.

To support such a high data rate, we need to propose new physical layer design to further reduce implementation complexity, save energy, and improve spectral efficiency. This is particularly true when the peak throughput requirement is imposed on a resource constrained (limited processing power, storage, and energy supply etc.) device. Since channel coding is well-known to consume a substantial proportion of computational resources, it poses a bottleneck for extreme throughput. To this end, channel coding is one of the most relevant physical layer technologies in order to guarantee $1Tbp/s$ peak throughput for 6G.

Polar codes, defined by Ar{\i}kan in \cite{ArikanPolar}, are a class of linear block codes with the generator matrix $G_N$ of size $N$, defined by $G_N \triangleq F^{\otimes n}$,
in which $N=2^n$ and $F^{\otimes n}$ denotes the $n$-th Kronecker power of $F=[\begin{smallmatrix}
1 & 0  \\
1 & 1
\end{smallmatrix} ]$.
Successive cancellation (SC) is a basic decoding algorithm for polar codes.

\begin{figure}
\centering
\includegraphics[width=0.4\textwidth]{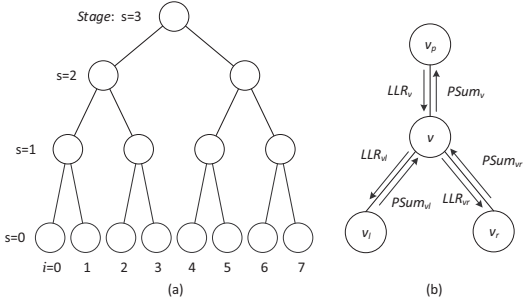} %
\caption{(a) decoding architecture as a binary tree; (b) node $\textit{v}$ received/response information}
\label{tree}
\end{figure}

Although the SC decoding algorithm seems unsuitable for high-throughput applications due to its serial nature, state-of-the-art SC decoders \cite{SSC}\cite{A1}\cite{Jiajie}\cite{A2}\cite{Polaran} managed to significantly simplify and parallelize the decoding process such that the area efficiency of SC decoding has far exceeded that of belief propagation (BP) decoding for low-density parity-check codes (LDPC). In particular, these works represent SC decoding as a binary tree traversal \cite{SSC}, as shown in Fig.~\ref{tree}(a). Each subtree therein represents a shorter polar code. The original SC decoding algorithm traverses the tree by visiting all the nodes and edges, leading to high decoding latency. Simplified SC decoders can fast decode certain subtrees (shorter polar codes) and thus ``prune'' those subtrees. The resulting decoding latency is largely determined by the number of remaining edges and nodes in the pruned binary tree. Several tree-pruning techniques have been proposed in \cite{SSC}, \cite{FSSCL} and \cite{FSSC}. To achieve $1Tbp/s$ throughput, more aggressive techniques need to be proposed on both the decoding and encoding sides.

\subsection{Contributions}
This paper introduces a novel polar code construction method, coined as ``fast polar codes'', to facilitate parallelized processing at an SC decoder. In contrast to some existing decoding-only techniques, we take a joint encoding-decoding optimization approach. Similar to existing methods, our main ideas could be better understood from the binary tree traversal perspective. They are (a) pruning more subtrees, (b) replacing some non-prunable subtrees with other fast-decodable short codes of the same code rates and then prune these ``grafted'' subtrees, (c) eliminating the remaining non-prunable subtrees by altering their code rates. As seen, both (b) and (c) involve a modified code construction. Consequently, we are able to fast decode any subtree (short code) of a certain size, without sacrificing parallelism.

The algorithmic contributions are summarized below:
\begin{enumerate}
    \item We introduce four new fast decoding modules for nodes with code rates $\{\frac{2}{M},\frac{3}{M},\frac{M-3}{M},\frac{M-2}{M}\}$. Here $M = 2^s$ is the number of leaf nodes in a subtree, where $s$ is the stage number. These nodes are called dual-REP (REP-2), repeated parity check (RPC), parity checked repetition (PCR), dual-SPC (SPC-2) nodes, respectively. More importantly, these modules reuse existing decoding circuits for repetition (REP) and single parity check (SPC) nodes.
    \item For medium-code-rate nodes that do not natively support fast decoding, we graft two extended BCH codes to replace the original outer polar codes. BCH codes enjoy good minimum distance and natively support efficient hard-input decoding algorithms, thus strike a good balance between performance and latency. The extension method is also customized to enhance performance.
    \item We propose to re-allocate the code rates globally, such that all nodes up to a certain size support the above mentioned fast decoding algorithms. This approach completely avoids the traversal into certain ``slow'' nodes.
\end{enumerate}

For code length $N = 1024$ and code rate $R = 0.875$, the proposed fast polar codes enable parallel decoding of all length-$16$ nodes. The proposed decoding algorithm reduces $55\%$ node visits and $43.5\%$ edge visits from the original polar codes, with a cost of within $0.3dB$ performance loss. Two types of decoder hardware are designed to evaluate the area efficiency and energy efficiency.

The implementation-wise contributions are summarized below:
\begin{enumerate}
    \item We design a recursive decoder to flexibly support any code rates and code lengths $N \leq 1024$. This decoder layout area is only $0.045mm^2$. For code length $N = 1024$ and code rate $R = 0.875$, it achieves a $25.6Gbp/s$ code bit throughput, with an area efficiency of $561Gbps/mm^2$.
    \item We also design an unrolled decoder that only supports one code rate and code length. The decoder layout area is $0.3mm^2$. For code length $N = 1024$ and code rate $R = 0.875$, it provides a $1229Gbp/s$ code bit throughput, with an area efficiency of $4096Gbps/mm^2$.
\end{enumerate}

\section{From simplified SC decoding to fast polar codes}\label{new cons}

Following the notations in \cite{SSC}, a node $v$ in a tree is directly connected to a parent node $p_{v}$, left child node $v_{l}$ and right child node $v_{r}$, respectively\footnote{A leaf node $v_{leaf}$ has no child node, and a root node $v_{root}$ has no parent node.}.
The stage of a node $v$ is defined by the number of edges between node $v$ and its nearest leaf node. All leaf nodes are at stage $s=0$.
The set of nodes of the subtree rooted at node $\textit{v}$ is denoted by $V_{v}$. Thus $V_{root}$ denotes the full binary decoding tree.
The set of all leaf nodes is denoted by $U$, the index of a leaf $\textit{u}$ \cite{SSC} is denoted by $l(u)$, and the indices of $U$ is denoted by $l(U)$.
Meanwhile, the set of the leaf nodes in subtree $V_{v}$ is denoted by $U_v$, and the indices of $U_v$ is denoted by $l(U_v)$.

The set of all information bit positions is denoted by $\mathcal{I}$ and that of all frozen bits by $\mathcal{I}^{c}$. The set of the information bit positions in subtree $V_{v}$ is denoted by $\mathcal{I}_{v}$ and the remaining frozen bit positions therein by $\mathcal{I}^{c}_{v}$.

\subsection{Simplified SC Decoding}\label{bch}
If $\mathcal{I}^{c}_{v}$ matches patterns, a so-called pattern-based simplified decoding can be triggered to process the node in parallel rather than bit-by-bit. From the binary tree traversal perspective, all the child nodes of $v$ do not need to be traversed. Thus decoding latency is reduced.

The existing so-called pattern-based simplified decoding includes $4$ different types. A node $v$ is a Rate-1 node\cite{SSC} if all leaves in the subtree $V_{v}$ are information bits, and a Rate-0 node\cite{SSC} if all leaves in the subtree $V_{v}$ are frozen bits. To improve the decoder's efficiency, \cite{FSSCL} defines single parity check (SPC) and repetition (REP) nodes.
We can employ pattern-specific parallel processing for each type of nodes.
Obviously, we need to identify and exploit more special nodes or patterns for latency reduction.

In this paper, we present four new types of corresponding nodes:
\begin{itemize}
  \item Define a node $v$ as a dual-SPC (SPC-2) node if $V_v$ includes only two frozen bits, and the frozen bits indices are the two smallest in $l(U_v)$.
  \item Define a node $v$ as a dual-REP (REP-2) node if $V_v$ includes only two information bits, and the information bits indices are the two largest in the $l(U_v)$.
  \item Define a node $v$ as a repeated parity check (RPC) node if $V_v$ includes only three frozen bits, and the frozen bits indices are the three smallest in the $l(U_v)$.
  \item Define a node $v$ as parity checked repetition (PCR) node if $V_v$ includes only three information bits, and the information bits indices are the three largest in the $l(U_v)$.
\end{itemize}

We describe their corresponding fast decoding methods in Section \ref{sec:SPNODE}.

Pattern-based simplified decoding skips the traversal of certain subtrees when it matches the above patterns.

Currently, there are eight pattern types to cover eight code rates of a sub tree: $\{0,\frac{1}{M},\frac{2}{M},\frac{3}{M},\frac{M-3}{M},\frac{M-2}{M},\frac{M-1}{M},1\}$. In other words, nodes with other code rates cannot be fast decoded. We need to work on the following two parameters.
\begin{enumerate}
  \item Ratio of simplified nodes: currently eight out of the $M+1$ code rates support simplified decoding. The ratio is thus $\frac{8}{M+1}$. Note that only the lowest and highest codes rates can be simplified, meaning code rates between $\frac{3}{M}$ and $\frac{M-3}{M}$ do not benefit from the fast decoding algorithm. For short and medium length codes, many nodes fall into this range due to insufficient polarization. We hope to further reduce latency by introducing more fast-decodable patterns to cover more code rates.
  \item Degree of parallelism: it can be represented by $M$, since the $M$ bits in a simplified node are decoded in parallel. The larger $M$ is, a larger proportion of the binary tree can be pruned due to simplified decoding. we hope to increase $M$ for higher throughput as well.
\end{enumerate}

For $M=8$, the ratio of simplified nodes is $8/9$, with only one code rate $\frac{4}{8}$ unsupported, but the degree of parallelism is only $8$. For $M=16$, the ratio of simplified nodes reduces to $8/17$, leaving a wide gap of nine unsupported code rates $\frac{4}{16},...,\frac{12}{16}$, but the degree of parallelism doubles.

\subsection{BCH node}\label{bch}

To cover medium code rates, we need to find some patterns which can be fast decoded with good BLER performance.
The bad news is, to the best of our knowledge, there exists no parallel decoding method for polar codes with code rates between $\frac{3}{M}$ and $\frac{M-3}{M}$. The good news is that the outer codes represented by a subtree can be replaced by any codes, as shown in many previous works \cite{Ying-14} \cite{Hamid-17} \cite{Dina-19}.
A good solution is removing the polar nodes with code rate falling into the gap, and grafting a different code that allows fast decoding.

BCH codes are good candidates due to their good minimum-distance property and fast hard-input decoding algorithms. If the error correcting capability is $t$, it is easy to design BCH codes whose minimum Hamming distance is larger than $2\times t$. This leads to good BLER performance.
Meanwhile, the Berlekamp-Massey (BM) algorithm can decode a BCH code with $t=1$ or $t=2$ within a few clock cycles. When grafted to polar codes as fast-decodable nodes, hard decisions are applied to the LLRs from the inner polar codes (parent nodes) before sending to the outer BCH codes (child nodes). Here the BCH codes are called ``BCH nodes''.

But BCH codes do not readily solve our problem. They only support a few code rates and code lengths, meaning they cannot cover all the codes rates within the gap.
For the degree of parallelism $M=16$, the target code length is $2^4$, so the nearest code length of BCH is $15$. Meanwhile, BCH codes only support code rates $\frac{7}{15}$ and $\frac{11}{15}$ within the gap and the corresponding number of information bits are $k=7, k=11$.

To overcome the issues, we first extend the code length to $16$ bits. For the BCH codes with $k=7$ and $t=2$, the original codes can correct two error bits. We add an additional bit to be the parity check of all BCH code bits.
The proposed two-step hard decoding works as follows. When the hard decision incurs three bit errors, and one of the errors has the minimum amplitude, the SPC bit can help correct one error bit first. Then the remaining two error bits can be corrected by the BM algorithm. But the same SPC extension no longer works for BCH codes with $k=11$ and $t=1$. The reason is as follows. If there are two or more bit errors in the node, the SPC function and BM algorithm both fail. Else if there is one error, the failure of SPC decoding will lead to more errors during BM decoding. Instead of SPC extension, we repeat one BCH code bit to improve its reliability.

Now that we have grafted two types of BCH nodes, the pattern-based decoding can support 10 code rates. The ratio of simplified nodes increases to $10/17$, and the maximum gap reduces to $\frac{4}{16}$. Figure \ref{node} shows the code rates supported by pattern-based decoding for degree of parallelism $M=16$.
\begin{figure}
\centering
\includegraphics[width=0.5\textwidth]{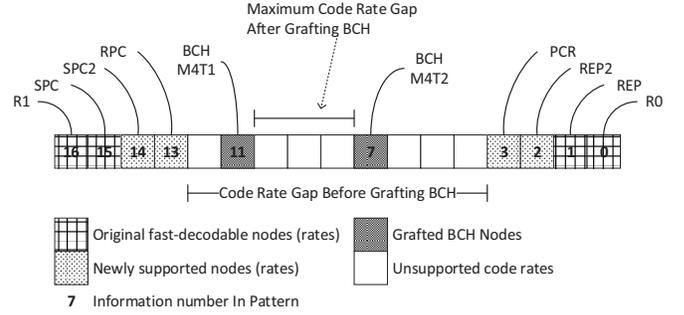} %
\caption{Nodes (code rates) supporting fast decoding for degree of parallelism $M=16$.}
\label{node}
\end{figure}

\subsection{Fast polar codes via rate re-allocation}\label{bch}
Even with the inclusion of BCH nodes, the fast decoding algorithm could not cover all the code rates of length-16 subtrees. As the second part of the solution, we propose to construct  fast polar codes to avoid the ``slow'' nodes, and only use the existing ten patterns. Here ``fast'' resembles that of fast SC decoding but is achieved by altering the code construction instead of decoding. We show that it greatly reduces decoding latency and increases throughput with only slight performance loss.

The following steps show how to construct fast polar codes only with node patterns of discontinuous code rates:
\begin{enumerate}
  \item Employ traditional methods such as Gaussian approximation (GA) or polarization weight (PW) to build polar codes with the parameter of code length $N$ and code rate $R$.
  \item Split all $N$ synthesized sub-channels to $N/16$ segments. Each segment constitutes a $16$-bit long block code, or equivalently a subtree with $16$ leaf nodes.
  \item Find out all ``slow'' segments which do not match the supported code rates or patterns. Re-allocate the code rates among segments to match the nearest supported code rate or pattern, which has $K$ information bits.
  \item If the number of information bits of the current segment exceeds or fall short of $K$, we remove or add a few information bits according to reliability. Apply this process to the remaining ``slow'' segments until all segments become fast-decodable.
\end{enumerate}

The resulting code is coined as ``fast polar code''. A detailed description of the construction algorithm for fast polar codes can be found in Appendix \ref{FP}.

Take code length $N=1024$, code rate $R=0.875$ as an example, we count the number of fast-decodable nodes to be visited, $f_{+/-}$-functions \cite{Dec:LLR_based_SCL} to be executed and edges to be traversed. These numbers provide a good estimate of SC decoding latency \cite{SSC} \cite{FSSCL}, and are thus used to compare between the construction proposed in this section and the GA construction in Table \ref{Comparison}. As seen, the traversed nodes and edges reduce by $55\%$ and $43.5\%$ , respectively, while the $f_{+/-}$-function executions reduce only by $8.9\%$. Note that the former two parameters have a greater influence than $f_{+/-}$-functions because it cannot be parallelized in any form.
\begin{table}
  \renewcommand{\arraystretch}{1.15}
    \caption{Comparison of traversed nodes, edges and executed $f_{+/-}$ Between GA construction and the proposed Fast Polar code Construction}
    \label{Comparison}
    \centering
    \begin{tabular}{|c|c|c|c|c|c|c|c|}
    \hline
    \multicolumn{8}{|c|}{Distribution of fast-decodable nodes}\\
    \hline
      \multicolumn{4}{|c|}{GA Construction} & \multicolumn{4}{c|}{Fast Polar Code Construction} \\
    \cline{1-8}
       Rate-1    & 4  & SPC    &20   &Rate-1    & 2  & SPC    &9\\
    \cline{1-8}
       SPC-2 & 2  & RPC    &0   &SPC-2 & 1  & RPC  &0 \\
    \cline{1-8}
       PCR   & 1  & REP-2  &1   &PCR   & 3  & REP-2  &1 \\
    \cline{1-8}
       REP   & 11  & Rate-0     &1   &REP   & 1  & Rate-0     &1 \\
    \cline{1-8}
       BCH t=1  & 0  & BCH t=2   &0   &BCH t=1  & 3  & BCH t=2   &2 \\
    \hline
      \multicolumn{8}{|c|}{Count with respect to binary tree traversal}  \\
    \hline
      \multicolumn{1}{|c|}{ }     & \multicolumn{2}{c|}{GA} & \multicolumn{2}{c|}{Fast} & \multicolumn{3}{c|}{Reduction(\%)} \\
    \hline
      \multicolumn{1}{|c|}{Nodes} & \multicolumn{2}{c|}{40} & \multicolumn{2}{c|}{22}   & \multicolumn{3}{c|}{55\%}\\
    \hline
      \multicolumn{1}{|c|}{$f_{+/-}$}& \multicolumn{2}{c|}{4160} & \multicolumn{2}{c|}{3792} & \multicolumn{3}{c|}{8.9\%}\\
    \hline
      \multicolumn{1}{|c|}{edges} & \multicolumn{2}{c|}{76} & \multicolumn{2}{c|}{43} & \multicolumn{3}{c|}{43.5\%}\\
    \hline
    \end{tabular}
\end{table}

\begin{figure}
\centering
\includegraphics[width=0.5\textwidth]{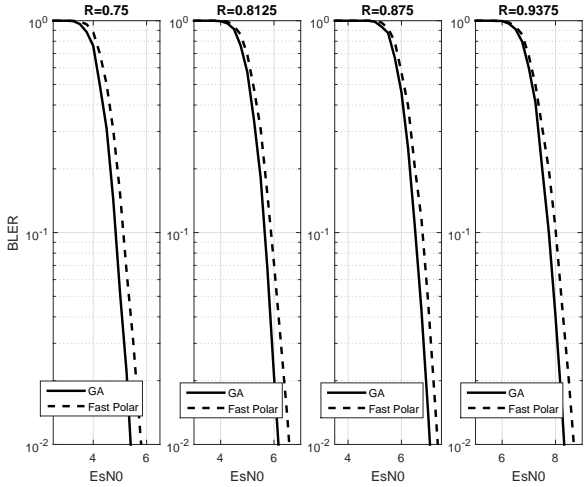} %
\caption{BLER Performance comparison between GA and fast polar code construction.}
\label{node}
\end{figure}
It is worth noting that the proposed fast polar code construction algorithm reallocates the code rates of some nodes against their actual capacity derived from channel polarization. This inevitably incurs BLER performance loss. To evaluate the loss, we run simulations and Fig. \ref{node} compares the BLER curves of both constructions under code length $N=1024$, and code rates $R=\{0.75,0.8125,0.875,0.9375\}$. There is a maximum of 0.3dB loss at BLER $10^{-2}$ between GA polar codes and the fast polar codes when adopting QPSK modulation.

\section{Fast decoding algorithms}\label{sec:SPNODE}

In this section, we describe the algorithms to support fast decoding of the newly defined SPC-2, REP-2, RPC and PCR nodes. For BCH nodes, we employ the classic BM algorithm which takes hard inputs and supports hardware-friendly fast decoding.

Each fast-decodable node $v$ at stage $s$ can be viewed as an outer code of length $M=2^s$. The code bits of $v$ as an outer code are denoted by $X_v$, with $M$ bits.
\subsection{SPC-2}\label{sec:SPC2}
For a dual-SPC node $v$, we divide its code bits $X_v$ into two groups, $X_v^{even}$ whose indices are even numbers, and $X_v^{odd}$ whose indices are odd numbers.
According to the definition of an SPC-2 node, there are two parity-check bits in the subtree $V_v$, and the corresponding parity functions $p[0]$ and $p[1]$ can be written as
\begin{equation*}\label{alp_cal}
    \left\{
        \begin{array}{lr}
        p[0]:\bigoplus x = 0,x\in X_v \\
        p[1]:\bigoplus x = 0,x\in X_v^{odd}
        \end{array}
    \right.
\end{equation*}
We add the two parity functions to get a parity function $p[2]$:
\begin{equation*}
p[2]=p[0]\oplus p[1]: \bigoplus x=0,x\in X_v^{even}
\end{equation*}
Since the two parity functions $p[1]$ and $p[2]$ involve two disjoint sets of code bits, the decoding of an SPC-2 node can be parallelized to two SPC nodes. Each SPC node inherits half of  the elements from $X_v$. We can reuse two SPC decoding modules to fast decode an the SPC-2 node.

\subsection{REP-2}\label{sec:REP2}
For a dual-REP node $v$, we divide its code bits $X_v$ into two groups, $X_v^{even}$ whose indices are even numbers, and $X_v^{odd}$ whose indices are odd numbers.
According to the definition of an REP-2 node, there are two information bits in the subtree $V_v$. They are denoted by $u_{M-2}$ and $u_{M-1}$.

It can be easily verified that $X_v^{odd}$ are the repetition of $u_{M-1}$ and $X_v^{even}$ are the repetition of $u_{M-2}\oplus u_{M-1}$. Accordingly, we can divide a length-$M$ dual-REP node into two $M/2$ REP nodes. We can reuse two REP decoding modules in parallel to fast decode a the REP-2 node.

\subsection{RPC}\label{sec:SPC3}
For an RPC node $v$, we divide its code bits $X_v$ into four groups as follows:
\begin{equation}\label{equ:xgroup}
X_v^{i}=\{x\in X_v, mod(l(x),4)=i\}, i\in\{0,1,2,3\}
\end{equation}

According to the definition of an RPC node, there are three parity-check bits in the subtree $V_v$, and the parity functions $p[0]$, $p[1]$ and $p[2]$ can be written as
\begin{equation*}\label{alp_cal}
    \left\{
        \begin{array}{lr}
        p[0]:\bigoplus x=0,x\in X_v^{0} \cup X_v^{1} \cup X_v^{2} \cup X_v^{3} \\
        p[1]:\bigoplus x=0,x\in X_v^{1} \cup X_v^{3} \\
        p[2]:\bigoplus x=0,x\in X_v^{2} \cup X_v^{3}
        \end{array}
    \right.
\end{equation*}
We add the latter two parity functions to get parity function $p[3]$:
\begin{equation*}
p[3]=p[1]\oplus p[2]: \bigoplus x=0,x\in X_v^{1} \cup X_v^{2}
\end{equation*}
And add this parity function to the first one to get parity function $p[4]$:
\begin{equation*}
p[4]=p[0]\oplus p[3]: \bigoplus x=0,x\in X_v^{0} \cup X_v^{3}
\end{equation*}
We define $\hat{c}_i=\bigoplus x,x\in X_v^{i}, i\in[0,1,2,3]$.
According to parity functions $p[1]$ to $p[4]$, one can easily verify that the following relationship holds:
\begin{equation}\label{check}
\hat{c}_1 \oplus \hat{c}_3 = \hat{c}_2 \oplus \hat{c}_3 = \hat{c}_1 \oplus \hat{c}_2 = \hat{c}_0 \oplus \hat{c}_3 = 0
\end{equation}
Equation \eqref{check} implies the existence of a virtual repetition code of rate $\frac 1 4$, because:
\begin{equation*}
\hat{c}_0 = \hat{c}_1  = \hat{c}_2 = \hat{c}_3 = 0
\end{equation*}
or
\begin{equation*}
\hat{c}_0 = \hat{c}_1  = \hat{c}_2 = \hat{c}_3 =1
\end{equation*}
where $\hat c_0, \hat c_1, \hat c_2, \hat c_3, $ are the virtual repeated code bits.

Given the above knowledge, the decoding algorithm for an RPC node at stage $s$ where $s\leq2$, can be easily derived as Algorithm \ref{alg:spc3}, in which
$sig(\alpha)\triangleq \left\{
	\begin{array}{lcl}
    0, \alpha \geq 0\\
	1, \alpha < 0\\
	\end{array} \right. $.

\begin{algorithm}[htb]
	\caption{Decoding a repeated parity check (RPC) node.}
	\label{alg:spc3}
	\begin{algorithmic}[1]
		\REQUIRE ~~\\
		The received signal $\alpha_v = \{\alpha_{v_k}, k=0\cdots M-1\}$;\\
		\ENSURE ~~\\
        The codeword to be recovered: $\hat{\mathbf{x}} = \{\hat{x_k}, k=0\cdots M-1\}$;\\
        \STATE Initialize: $\Delta_0 = 0, \Delta_1 = 0$ \\		
        \STATE Initialize: $\delta_i = \infty, c_i = 0, p_i = 0$ for $i=0\cdots 3$; \\
        \STATE Initialize: $\hat{x_k} = sig(\alpha_{v_k})$ for $k=0\cdots M-1$; \\
		\FOR {$i=0 \cdots 3$}
		\FOR {$j=0 \cdots M/4$}		
		\STATE $k=j\times 4+i$;
		\STATE $c_i = c_i \oplus sig(\alpha_{v_k})$;
        \STATE if $\left |\alpha_{v_k}\right | < \delta_i$
        \STATE \quad $p_i = k$;
        \STATE \quad $\delta_i = \left |\alpha_{v_k}\right |$;
		\ENDFOR
        \STATE if $c_i = 1$
        \STATE \quad $\Delta_0 = \Delta_0 + \delta_i$
        \STATE else
        \STATE \quad $\Delta_1 = \Delta_1 + \delta_i$
		\ENDFOR
		\FOR{$i=0\cdots 3$}
		\STATE if$((\Delta_0 > \Delta_1) \cap (c_i = 0)) \cup ((\Delta_0 < \Delta_1) \cap (c_i = 1)) $
        \STATE \quad $\hat{x_{p_{i}}} = \sim\hat{x_{p_{i}}}$
		\ENDFOR
	\end{algorithmic}
\end{algorithm}

\subsection{PCR}\label{sec:REP2}
For a PCR node $v$, we divide its code bits $X_v$ into four groups in the same way as in \eqref{equ:xgroup}.
According to the definition of an RPC node, there are three information bits in this node. They are denoted by $u_{M-3}$, $u_{M-2}$ and $u_{M-1}$.

We define $c_i, i\in \{0,1,2,3\}$ according to the following equation
\begin{equation}\label{equ:k}
[c_0~c_1~c_2~c_3] = [0~u_{M-3}~u_{M-2}~u_{M-1}] \times G_4
\end{equation}

It can be easily verified that $X_v^{0}$ are the repetition of $c_0$, $X_v^{1}$ are the repetition of $c_1$, $X_v^{2}$ are the repetition of $c_2$, and $X_v^{3}$ are the repetition of $c_3$. Thus, we divide the input signal $\alpha_v$ into four groups according the indices and combine the input signals within each group into four enhanced signals $\Delta_i, i\in \{0,1,2,3\} $, as in an REP node.

Equation \eqref{equ:k} implies the existence of a virtual single parity check code of rate $\frac 3 4$, with virtual code bits $c_i, i\in \{0,1,2,3\}$, so we can reuse SPC module to decode it. A detailed description of PCR decoding is given in Algorithm \ref{alg:rep23}.

\begin{algorithm}[htb]
	\caption{Decoding a parity checked repetition (PCR) node.}
	\label{alg:rep23}
	\begin{algorithmic}[1]
		\REQUIRE ~~\\
		The received signal $\alpha_v = \{\alpha_{v_k}, k=0\cdots N-1\}$;\\
		\ENSURE ~~\\
        The codeword to be recovered: $\hat{\mathbf{x}} = \{\hat{x_k}, k=0\cdots N-1\}$;\\
        \STATE Initialize: $\Delta_i = 0$ for $i=0\cdots 3$; \\		
		\FOR {$i=0 \cdots 3$}
		\FOR {$j=0 \cdots N/4$}		
		\STATE $k=j\times 4+i$
        \STATE $\Delta_i = \Delta_i + \alpha_{v_k}$
		\ENDFOR
		\ENDFOR
        \STATE $\{\hat{c_0},\hat{c_1},\hat{c_2},\hat{c_3}\}$ = SPC\_DEC($\{\Delta_0,\Delta_1,\Delta_2,\Delta_3\}$)
		\FOR{$i=0\cdots 3$}
        \FOR {$j=0 \cdots N/4$}		
        \STATE $k=j\times 4+i$
        \STATE $\hat{x_k} = \hat{c_i}$
		\ENDFOR
        \ENDFOR
	\end{algorithmic}
\end{algorithm}

\section{Hardware Implementation}\label{hardware}
We designed two types of hardware architectures to verify the performance, area efficiency and energy efficiency.
\begin{itemize}
\item \textbf{\emph{Recursive Decoder}}: It supports flexible code length and coding rates of mother code length $N$  from $32$ to $1024$ with the power of $2$. With rate matching, flexible code length with $0 < N \leq 1024$ and code rate with $0 < R \leq 1$ are supported. The $f_{+/-}$ functions in nodes are processed by single PE (processing element) logic, and one decision module to support all 9 patterns\footnote{R0 node is bypassed in SC decoding.}. The decoder processes one packet at a time.
\item \textbf{\emph{Unrolled Decoder}}: It only supports a fixed code length and code rate. In our architecture we hard coded code length $N=1024$, and code rate $R=0.875$.
This fully unrolled pipelined design combines exclusive dedicated PEs to process each $f_{+/-}$ function in the binary tree.
Same to the decision modules that $21$ dedicated node specific logic are implemented to support $21$ nodes patterns.
With $25$ packets simultaneously decoding, thanks to the unrolled fully utilization of processing logic and storage, this decoder provides extreme high throughput with high area efficiency and low decoding energy.
\end{itemize}

Both the above mentioned decoder implementations adopt successive cancellation algorithm accelerated by pattern-based fast decoding. The maximum degrees of parallelization are $128$ for SPC and SPC-2 nodes, and $256$ for R1 nodes. All other nodes enjoy a degree of parallelism of $16$.

\subsection{Parallel Comparison Circuit}\label{recu}
We observe that there are several large SPC nodes in the right half of the binary tree. As described, these SPC nodes need to be processed with a higher degree of parallelism to achieve a higher throughput. The SPC decoding algorithm is very simple as follows. First, get the signs of an SPC node's input signals, find the minimum amplitude of input signals and record its position. Then, do a parity check of the signs. If it passes, then return these signs, else reverse the sign of recorded minimum-amplitude position and return the updated signs.

To process a large SPC node, a circuit is required to locate a minimum amplitude from a large amount of input signals. The traditional pairwise comparison method requires a circuit of depth $log_2(M)$, where $M$ is the number of amplitudes to be compared.
Finding the smallest among eg., $128$ amplitudes takes $7$ steps comparison, considering clock frequency is at $1Ghz$, it is very challenging to meet timing constraints completing all comparisons in one clock cycle.

We advocate a parallel comparison architecture to replace the traditional one.
For a node $v$ at stage $s$, its input signals $\alpha_{v}$ include $M=2^s$ elements, the amplitudes of which are denoted as $\left [A_{0}~A_{1}~\cdots~A_{M-1}\right ]$. Each amplitude has $x$-bit quantization. We fill the $x$-bit quantized binary vectors into the columns of a matrix as follows:
\begin{equation*}
\left[A_{0}~\cdots~A_{i}~\cdots~A_{M-1}\right] = \begin{bmatrix} b_{0}^{0} & \cdots & b_{i}^{0} & \cdots & b_{M-1}^{0} \\
                                                 \vdots    & \ddots & \vdots    & \ddots & \vdots  \\
                                                 b_{0}^{j} & \cdots & b_{i}^{j} & \cdots & b_{M-1}^{j} \\
                                                 \vdots    & \ddots & \vdots    & \ddots & \vdots  \\
                                                 b_{0}^{x-1} & \cdots & b_{i}^{x-1} & \cdots & b_{M-1}^{x-1} \end{bmatrix}
\end{equation*}
Rewrite the matrix with respect to its row vectors matrix and we have $\left[B_{0}~\cdots~B_{j}~\cdots~B_{x-1}\right]^\mathsf{ T }$, in which $B_{j}=\left[b_{0}^{j}~\cdots~b_{i}^{j}~\cdots~b_{M-1}^{j}\right ], j \in\{0,1\cdots x-1\}$ is a row vector. $B_{j}$ can be represented as an $M$-bit variable. We propose Algorithm \ref{alg:spc} to find out the minimum-amplitude position through a reverse mask $D$, in which the bit ``1'' indicates the minimum.

\begin{algorithm}[htb]
	\caption{Parallel Comparison Algorithm.}
	\label{alg:spc}
	\begin{algorithmic}[1]
		\REQUIRE ~~\\
		The received signal $\alpha_v = \{\alpha_{v_k}, k=0\cdots M-1\}$;\\
		\ENSURE ~~\\
        The Reverse Mask: $D$ is an $M$-bit Variable;\\
        \STATE Initialize: $\left[B_{0}~\cdots~B_{j}~\cdots~B_{x-1}\right]^\mathsf{ T }$ from $\alpha_v$; \\
        \STATE Initialize: An $N$-bits variable $C=\bf{0}$, .
        \FOR {$j=x-1 \cdots 0$}	
        \STATE $M$-bit Variable $E = (C | B_{j})$
        \STATE if(Not all bits in $E$ are ``1'')
        \STATE \quad $C = E$
        \ENDFOR
        \STATE Reverse Mask $D=\sim C$
	\end{algorithmic}
\end{algorithm}

The parallel comparison algorithm reduces the comparison logic depth from $log_2(M)$ to 1.
But the reverse mask $D$ may have two or more minimum positions. That means the input signals $\alpha_{v}$ include two or more minimum amplitudes. It must generate an error if there are two minimum amplitudes. To avoid this error occur, we can apply an additional circuit to ensure the uniqueness of the selected minimum position.

\subsection{Bit quantization}\label{unroll}
An attractive property of polar codes is that SC decoding works well under low-precision quantization (4 bits to 6 bits). Lower precision quantization is the key to higher throughput, as it effectively reduces implementation area and increases clock frequency.

\begin{figure}
\centering
\includegraphics[width=0.45\textwidth]{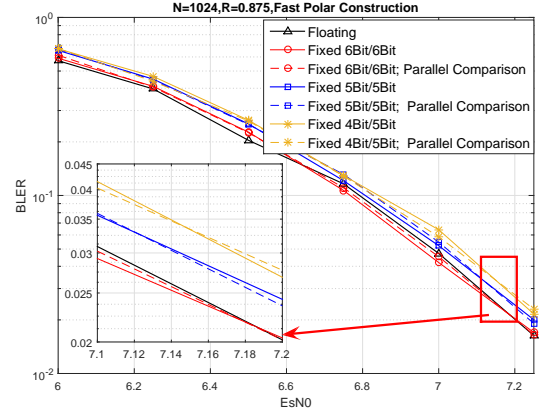} %
\caption{Performance comparison between Floating Point and Fixed Point.}
\label{N1024K896}
\end{figure}

There are two types of quantization numbers, one is for channel LLR and the other is for internal LLR.
We first test the case with $6$-bit input quantization and $6$-bit internal quantization. According to Fig \ref{N1024K896}, this setting achieves the same performance as floating-point.
The second one is $5$-bit quantization/$5$-bit internal quantization. It incurs $< 0.1$dB loss. Finally, $4$-bit input quantization/5-bit internal quantization incurs $< 0.2$dB loss. In this paper, we evaluate the physical implementation result under $5$-bit quantization both input and internal signals to strike a good balance between complexity and throughput.

At the same time, we also compare the BLER performance between the original SPC and parallelized SPC. None of the quantization schemes yields harmful loss.

\subsection{Layout View}\label{layout}
We carry out the two physical implementations for both the recursive and unrolled architecture.

With TSMC 16nm technology, the recursive decoder synthesis area is $0.032mm^2$, the clock frequency is $1.00Ghz$. The decoder's layout size is $192\mu m\times 234\mu m= 0.045 mm^2$.
With the same ASIC technology node, the unrolled decoder synthesis area is $0.17mm^2$, the clock frequency is $1.20Ghz$. The decoder's layout size is $500\mu m\times 600\mu m= 0.3 mm^2$.
Figure \ref{layout} shows the two layout graphs of the decoders. Note that the area of the unrolled decoder is actually much larger than the recursive decoder.

\begin{figure}
\centering
\includegraphics[width=0.45\textwidth]{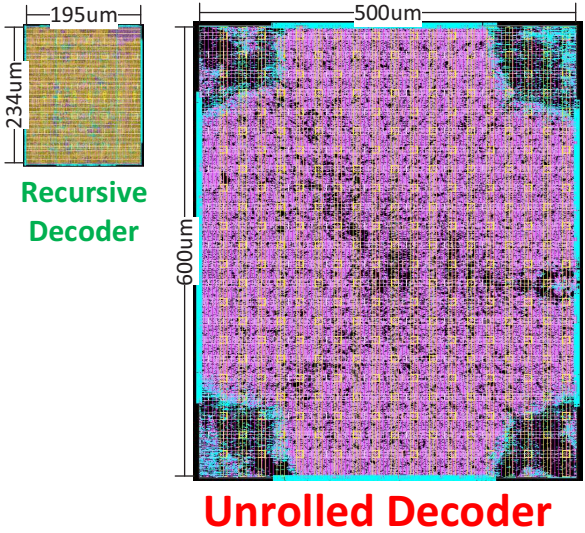} %
\caption{Layout Graph of Recursive and Unrolled Decoder under the same scale.}
\label{layout}
\end{figure}

\begin{table*}
  \renewcommand{\arraystretch}{1.05}
    \caption{COMPARISON WITH High Throughput POLAR DECODER}
    \label{table_comparsion}
    \centering
    \begin{tabular}{lccccc}
      Implementation    & This Work (Unroll) & This Work (Recursive) & \cite{Jiajie} & \cite{GN} & \cite{Polaran}\\
      \hline
      Construction      &Fast-Polar           &Fast-Polar            & Polar         & Product-Polar   &Polar \\
      Decoding Algorithm  & Fast-SC           & Fast-SC              & SC            & PDF-SC        &OPSC   \\
      Code Length       & 1024                & 1024                 & 32768         & 16384         &1024 \\
      Code Rate         &0.875                & 0.875                & 0.864         & 0.864         &0.83 \\
      \hline
      Technology        &\multicolumn{5}{c}{All in TSMC $16nm$} \\
      Clock Frequency($Ghz$)&1.20             &1.00                  &1.00           &1.05           &1.20  \\
      Throughtput/Coded-bit ($Gbps$)  &1229   &25.6                 &5.27           &139.7          &1229 \\
      Throughtput/Info-bit  ($Gbps$)  &1075   &22.4                 &4.56           &120.73         &1020 \\
      Area/Layout($mm^2$)    &0.30            &0.045                &0.35           &1.00           &0.79\\
      Area Eff/Coded-bit($Gbps/mm^2$) & 4096  &561                  &15.1           &139.7          &1555   	 \\
      Power($mW$)      &  784                 &30.9                  &-              &94             &1167 \\
      Energy($pJ/bit$) &  0.63                &1.21                  &-              &0.67           &0.95 \\
      \hline
    \end{tabular}
\end{table*}

\section{Key Performance Indicators}\label{KPI}

The key performance indicators (KPIs) are reported in this section. First of all, we evaluate the area efficiency using equation $Area Eff(Gbps/mm^{2}) = \frac{Info\,Size(bits)}{Latency(ns)\times Area(mm^{2})}$.

The recursive decoder takes 40 clock cycles to decoder one packet under fast polar code construction with code length $N=1024$, and code rate $R=0.875$. Thus the throughput is $(1024~bits\times 1~Ghz)/40 ~cycles= 25.6Gbps$ for coded bits, and $((1024\times 0.875) ~bits\times 1~Ghz)/40 ~cycles= 22.4Gbps$ for information bits. With TSMC $16nm$ process, the area efficiency for coded bits is $561Gbps/mm^2$.

The unrolled decoder takes $25$ clock cycles to decoder one packet. It is fully pipelined, meaning a new packet of decoded results would be generated continuously every cycle after the first $25$ clock cycles of the first packet processing time.
The throughput is thus $1024~bits \times 1.2~Ghz = 1229Gbps$ for coded bits, and $(1024\times 0.875) ~bits\times 1~Ghz = 1075Gbps$ for information bits. With TSMC $16nm$ process, the area efficiency for coded bits is $4096Gbps/mm^2$.

We further evaluate the power consumption and decoding energy per bit through a simulation in which 200 packets are decoded. The process, voltage and temperature (PVT) condition of evaluation is TT corner, $0.8V$ and $20^{\circ}C$, and the resulting of recursive decoder's power consumption is $30.9mW$, and decoding each bit costs $1.21pJ$ of energy on average; while the unrolled decoder's power consumption is $784mW$, and decoding each bit costs $0.63pJ$ of energy on average.

We also compare the decoding throughput, area efficiency and power consumption with several high-throughput decoders in literature, and present the results in Table \ref{table_comparsion}.
From the KPIs, we conclude that unrolled decoders are more suitable for scenarios requiring extremely high throughput but only support fixed code length and rate; recursive decoders are much smaller, which are better for resource constrained devices, and at the same time provides flexible code rates and lengths - a desirable property for wireless communications.

\section{Conclusions}
\label{section_conclusions}
In this paper, we propose a new construction of fast polar codes, which is solely composed of fast-decodable special nodes at length $16$. By viewing the decoding process as a binary tree traversal, the fast polar codes can reduce 55\% of node visits, $8.9\%$ of $f_{+/-}$ calculation and $43.5\%$ of edge traversal over the original polar construction at code length $N=1024$, and code rate $R=0.875$, at the cost of slight BLER performance loss.

We implement two types of decoders for the fast polar codes. The recursive decoder can support flexible code lengths and code rates, and support code length up to 1024. This decoder layout area is only $0.045mm^2$, and can provide $25.6Gbps$ coded bits throughput, with an area efficiency of $561Gbps/mm^2$.

The unrolled decoder only supports one code length $N=1024$ and one code rate $R=0.875$. However, the fully pipelined structure leads to hardware with ultra-high area efficiency and low decoding power consumption. This decoder layout area is $0.3mm^2$, and can provide $1229Gbps$ code bit throughput, with an area efficiency as high as $4096Gbps/mm^2$.

These results indicate that fast polar codes can meet the high-throughput demand in the next-generation wireless communication systems. And the recursive hardware design and unrolled hardware design can be adopted to satisfy different system requirements.

\appendix
\subsection{Fast Polar Code Construction Algorithm}\label{FP}
\begin{algorithm}[htb]
	\caption{A method to construct fast polar codes.}
	\label{alg:stage-permute-decoder}
	\begin{algorithmic}[1]
		\REQUIRE ~~\\
		    Code length $N$, information length $K$, the set of fast-decodable modes $\Theta$. \\
		\ENSURE ~~\\
		    Re-allocate node-wise code rates such that all nodes support fast decoding. \\
		\STATE Construct an $(N,K)$ polar code based on GA or PW methods.\\
    	\STATE Divide the code into segments of length $16$ and the number of segments is denoted by $N_s$.
        \STATE Progressively refine the code construction as follows. All the frozen bit positions are intialized as active states and ``active'' bit position can be transformed to an informtaion bit position in the refining process.
              \FOR {$t=1 \cdots N_s$}
       		       \WHILE {the $t$-th segment does not belong to $\Theta$}
                       \STATE Denote by $i$ the least reliable information bit position in the $t$-th segment.
                        \STATE Denote by $j$ the most reliable frozen bit position of active states in the subsequent segments, and denote by $k_j$ the number of information bits in that segment.
                        \IF  {$k_j \geq	11$ and $k_j<16$ or $k_j<3$}
                          \STATE Mark $i$ as a frozen bit position and $j$ as an information bit position.
                        \ELSE
                           \STATE Mark $j$ as inactive state.
                        \ENDIF
                    \ENDWHILE
 	          \ENDFOR
	\end{algorithmic}
\end{algorithm}

\end{document}